\documentclass[twocolumn,floatfix,superscriptaddress,amsmath,amsfonts,amssymb,aps]{revtex4}

\usepackage{graphicx}
\usepackage{dcolumn}
\usepackage{bm}
\usepackage[normalem]{ulem}
\usepackage{color}
\usepackage{amssymb}
\usepackage{amstext}
\usepackage{amsbsy}

\begin{document}

\preprint{APS/123-QED}

\title{Bismuth layer properties in the ultrathin Bi--FeNi 
multilayer films \\ probed by spectroscopic ellipsometry}

\author{N. N. Kovaleva}
\email{kovalevann@lebedev.ru}
\affiliation{Lebedev Physical Institute, Russian Academy of Sciences, Leninsky prospect 53, 119991 Moscow, Russia}
\author{D. Chvostova}
\affiliation{Institute of Physics, Academy of Sciences of the Czech Republic, Na Slovance 2, 18221 Prague, Czech Republic}
\author{O. Pacherova}
\affiliation{Institute of Physics, Academy of Sciences of the Czech Republic, Na Slovance 2, 18221 Prague, Czech Republic}
\author{A. V. Muratov}
\affiliation{Lebedev Physical Institute, Russian Academy of Sciences, Leninsky prospect 53, 119991 Moscow, Russia}
\author{L.~Fekete}
\affiliation{Institute of Physics, Academy of Sciences of the Czech Republic, Na Slovance 2, 18221 Prague, Czech Republic}
\author{I. A. Sherstnev}
\affiliation{Lebedev Physical Institute, Russian Academy of Sciences, Leninsky prospect 53, 119991 Moscow, Russia}
\author{K. I. Kugel}
\affiliation{Institute for Theoretical and Applied Electrodynamics, Russian Academy of Sciences, 125412 Moscow, Russia} 
\affiliation{National Research University Higher School of Economics, 101000 Moscow, Russia}
\author{F. A. Pudonin}
\affiliation{Lebedev Physical Institute, Russian Academy of Sciences, Leninsky prospect 53, 119991 Moscow, Russia}
\author{A. Dejneka}
\affiliation{Institute of Physics, Academy of Sciences of the Czech Republic, Na Slovance 2, 18221 Prague, Czech Republic}

\date{\today}

\begin{abstract}
Using wide-band (0.5--6.5\,eV) spectroscopic ellipsometry we 
study ultrathin [Bi(0.6--2.5\,nm)--FeNi(0.8,1.2\,nm)]$_{\rm N}$ 
multilayer films grown by rf sputtering deposition, where 
the FeNi layer has a nanoisland structure and its morphology 
and magnetic properties change with decreasing the nominal layer 
thickness. From the multilayer model simulations of the 
ellipsometric angles, $\Psi(\omega)$ and $\Delta(\omega)$, the 
complex (pseudo)dielectric function spectra of the Bi layer 
were extracted. The obtained results demonstrate that the Bi layer 
can possess the surface metallic conductivity, which is strongly 
affected by the morphology and magnetic properties of the nanoisland 
FeNi layer in the GMR-type Bi--FeNi multilayer structures.  

\end{abstract}

\pacs{Valid PACS appear here}

\maketitle
The spin-orbit coupling (SOC) is a relativistic effect important for 
the electronic structure of heavy atoms and solids formed by them. 
This leads to characteristic surface metallic states arising from 
the loss of the inversion symmetry at the surface (Rashba effect) 
\cite{Rashba}. 
Bismuth (Bi) is a rather heavy element with 
strong SOC in the atomic 6$p$ levels (where $p_{3/2}-p_{1/2}$ 
splitting is about 1.5 eV), which facilitates the application of 
quasi-two-dimensional (2D) Bi layers in spintronics as 
spin sources or filters, as well as in multilayer structures 
exhibiting the giant magnetoresistance (GMR) effect.
The scaling of Bi integrated units to smaller dimensions 
is still going on toward the thickness of 5\,nm and 
beyond, where 2D Bi (bismuthene) exhibits 
extraordinary electronic properties \cite{Liu_RSC}. 
For implementing the full potential of GMR applications by a
rational nanostructure design, the information on the electron 
band structure of 2D Bi layers is important. 

In bulk Bi, which crystallizes in the rhombohedral symmetry 
(space group $R{\bar 3}m$, unit cell parameters $a$\,=\,$b$\,=\,4.547\,$\AA$, $c$\,=\,11.8616\,$\AA$, $\alpha$\,=\,$\beta$\,=\,90$^\circ$, $\gamma$\,=\,120$^\circ$) with two atoms per unit cell, five bands accommodate ten valence 
electrons, which dictates an insulating behavior generically. 
However, the bands close to the Fermi level, namely, at the $T$ 
and $L$ points of the Brillouin zone, can be significantly 
affected by the strong SOC \cite{Gonze}. 
As a consequence, three conduction minima at the $L$ points 
lie at about 40\,meV lower than the single valence-band maximum 
at the $T$ point. This indirect band overlap implies 
the semimetallic behavior of bulk Bi with electron 
transport properties dictated by quite small electron 
effective mass along a certain axis and unusually long mean free 
path. Due to the crystal structure inversion 
symmetry, the SOC does not lead to any 
lifting of the spin degeneracy in the 6$p$ bands, each having two 
possible spin states per $k$ point in the Brillouin zone, while  
the loss of symmetry at the surface or interface can transform Bi 
from a semimetal (SM) to a metal when the electron or hole bands 
cross the Fermi level. 
\begin{figure}
\includegraphics*[width=75mm]{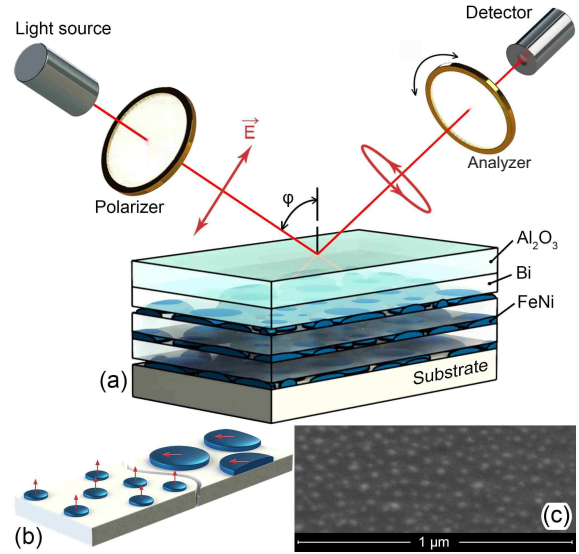}
\caption{Schematics of (a) the ultrathin [Bi--FeNi)]$_{\rm N}$ 
multilayer films investigated in the present study by spectroscopic ellipsometry and (b) the magnetization configuration in the nanoisland FeNi layer. 
Larger islands exhibit an in-plane magnetization configuration, 
while smaller islands can have an out-of-plane one 
\cite{Vedmedenko,Kovaleva_JNM} (for more details, see the text). 
(c) SEM image of the nanoisland FeNi layer on the Sitall substrate 
obtained by using a JEOL JSM-7001F facility. Reproduced with permission from Phys. Lett. A 410, 127546 (2021) \cite{Noskova}. 
Copyright 2021. Elsevier.\vspace{-0.5cm}}
\label{Scheme}
\end{figure}

In the thin-film limit, 
this effect can be at conflict with the quantum confinement effect 
(or size-effect), leading to complicated electronic properties. 
Quantum effects can be observed in thin films whose thickness 
is comparable to the effective wavelength of charge carriers, 
and their mean free path exceeds the film thickness. 
These conditions should transform Bi from a SM to 
a semiconductor (SC) at a critical film thickness of about 
300\,$\AA$ \cite{Hoffman}. Due to a very low charge carrier 
density ranging from about 10$^{17}$\,cm$^{-3}$ to 10$^{18}$\,cm$^{-3}$ 
and small relative effective masses of the charge carriers from 0.005 to 0.1, the optical excitation of charge carriers starts to be relevant 
only in the far infrared (below 0.1\,eV) 
\cite{Toudert1}, where a confinement-induced energy gap in thin Bi 
films could manifest itself in the optical experiments.     
However, the surface metallic states may hinder the SM-SC transition 
in ultrathin Bi films. The existence of the surface metallic states 
in ultrathin Bi(001) films was confirmed by the broadband terahertz 
time-domain spectroscopy study \cite{Yokota}. It was shown that the 
surface charge carrier density, plasma frequency, and 
scattering rate dramatically increase with a decrease 
in the film thickness, 
reaching $n$\,=\,3.1\,$\times$\,10$^{19}$\,cm$^{-3}$, $\omega_p$\,=\,4.0\,$\times$\,10$^2$\,THz (1.65\,eV), and
$\gamma_D$\,=\,4.8\,$\times$\,10$^2$\,THz (2.0\,eV), respectively, 
in the thinnest investigated 2.8\,nm Bi film \cite{Yokota}, 
where the estimated optical conductivity $dc$ limit  
$\sigma_{1(\omega\rightarrow 0)}=\omega_p^2/\gamma_D$\,
=\,2300\,$\Omega^{-1}\cdot$cm$^{-1}$. 

Recently, we have demonstrated that the electronic properties 
of the free and localized Ta charge carriers in (Ta--FeNi)$_{\rm N}$ 
multilayer films (MLFs) 
can be studied by spectroscopic 
ellipsometry (SE) \cite{Kovaleva_APL_2,Kovaleva_srep}.  
Here, we explore the elaborated SE approach to gain 
insights into the electron band structure and surface electronic 
properties of ultrathin Bi layers in real GMR-type  
(Bi--FeNi)$_{\rm N}$ MLF structures, 
incorporating nanoisland FeNi layers (see the scheme of the 
(Bi--FeNi)$_{\rm N}$ MLFs investigated in the present study by 
SE in Fig.\,\ref{Scheme}(a)). The morphology and magnetic properties 
of a single-layer nanoisland FeNi film grown on the Sitall substrate 
were studied earlier \cite{Kovaleva_JNM,Boltaev,Kovaleva_JNR}. 
Below the structural 
percolation transition at the nominal thickness of 1.5\,--\,1.8 nm 
\cite{Pudonin_JETP}, the FeNi layer is discontinuous and consists 
of inhomogeneously distributed FM nanoislands 
having lateral sizes of 5\,--\,30 nm and possessing giant magnetic 
moments of 10$^3$--10$^5$\,$\mu_{\rm B}$ (where $\mu_{\rm B}$ is 
the Bohr magneton). As an example, Fig.\,\ref{Scheme}(c) shows 
the scanning electron microscopy (SEM) image of the nanoisland FeNi 
film grown on the Sitall substrate \cite{Noskova}. 
As schematically shown in Fig.\,\ref{Scheme}(b), the larger islands 
(which appear closer to the percolation transition) have the in-plane 
magnetization configuration, while the smaller islands 
(existing quite far from the percolation transition) have 
the out-of-plane one \cite{Vedmedenko,Kovaleva_JNM}. 
Here, collective superferromagnetic (SFM) states -- FM or AFM -- 
may develop in the self-assembled local arrangements of 
FM nanoislands at comparatively high temperatures 
\cite{Kovaleva_JNM,Kleemann}. 
However, small and well separated FeNi nanoislands are weakly 
interacting via the magnetic dipole forces and exhibit 
superparamagnetic (SPM) behavior at high temperatures, 
which is associated with strongly fluctuating giant magnetic moments 
\cite{Kovaleva_JNM}. 
\begin{figure}
\includegraphics*[width=85mm]{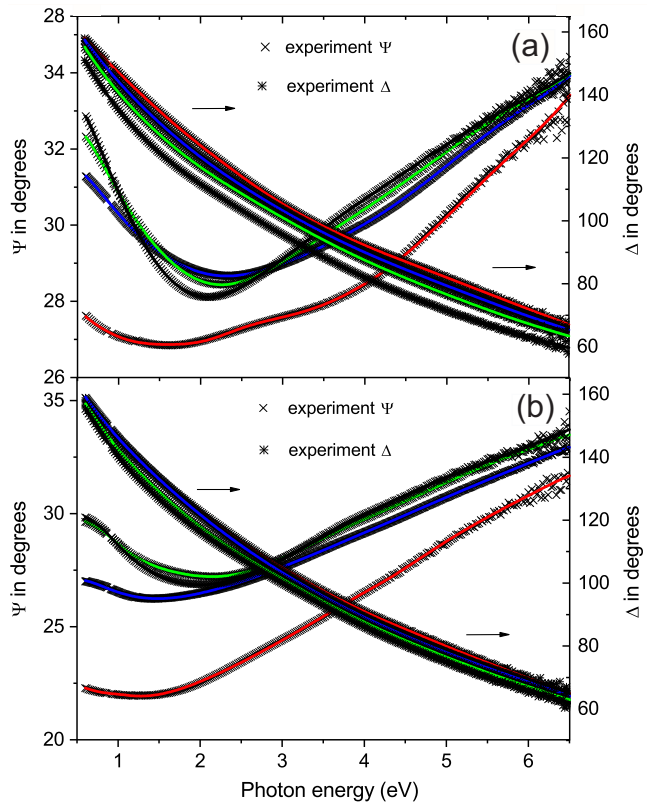}
\caption{Ellipsometric angles, $\Psi(\omega)$ and $\Delta(\omega)$, 
measured for the Al$_2$O$_3$(2.1\,nm)/[Bi--FeNi($h$)]$_{16}$/Sitall 
MLF samples having the FeNi layer thickness $h$ (a) 1.2\,nm and 
(b) 0.8\,nm at an angle of incidence of 70$^\circ$ (symbols) 
and the fitting results by the Drude-Lorentz model 
(Eq.\,(\ref{DispAna})) (displayed for the Bi layer thickness of 
2.5, 2.0, 1.4, and 0.6\,nm by solid black, green, blue, and red 
curves, respectively). 
\vspace{-0.5cm}}
\label{PsiDelta}
\end{figure}

The (Bi\,--\,FeNi)$_{\rm N}$ MLFs were grown 
by rf sputtering deposition from 99.95 \% pure Bi and 
Fe$_{21}$Ni$_{79}$ targets on glass Sitall (TiO$_2$) substrates. 
Before the deposition, the vacuum chamber was annealed at 200$^\circ$C, 
so that the prepared base pressure in the chamber was below  
2\,$\times$\,10$^{-6}$ Torr. During the deposition, the background Ar
pressure was 6$\times$10$^{-4}$ Torr. The actual temperature 
of the substrates was about 80$^\circ$C. We used the Sitall 
substrates with typical sizes of 15$\times$5$\times$0.6 mm$^3$. 
The Bi and FeNi layer nominal thickness was controlled by the 
deposition time determined by the film deposition rate. For example, 
the determined FeNi layer deposition rate was about 0.67\,$\AA$ 
per second. To protect the grown MLFs from the oxidation under ambient conditions, the as deposited MLFs were covered {\it in situ} by the 2.1\,nm-thick Al$_2$O$_3$ layer. In the prepared (Bi--FeNi)$_{\rm N}$ MLF samples,
the FeNi layer nominal thickness was 0.8 and 1.2\,nm, 
the thickness of the Bi layer was 0.6, 1.4, 2.0, and 2.5\,nm, 
and the number of Bi/FeNi bilayers was N\,=\,16. 
In our recent study, (Ta--FeNi) MLFs grown by rf sputtering deposition 
onto Sitall-glass substrates, including ultrathin 0.52\,nm-thick FeNi 
layers, were characterized by scanning/transmission electron 
microscopy (STEM) (for details see \cite{Kovaleva_srep}). 
Here, the grown Bi--FeNi MLF samples were 
characterized by the atomic-force microscopy (AFM), X-ray diffraction, 
and X-ray reflectivity (see supplementary material to this article).
The X-ray reflectivity measurements confirm a good periodicity and 
relatively small interface roughness in the grown Bi--FeNi MLF 
structures, as well as good agreement with the nominal thickness 
of the Bi and FeNi layers. The X-ray diffraction suggests 
orientation of the Bi layers along the (012) plane, 
where the interlayer distance is 3.28\,$\AA$. 
Thus, the prepared Bi--FeNi MLFs having 
the Bi layer thickness of 0.6, 1.4, 2.0, and 2.5\,nm 
correspond to about 2, 4, 6, and 8 Bi(012) monolayers.     
\begin{table}
\caption{\label{tab:table1}
Parameters of the Drude and low-energy Lorentz bands for the  
Bi layer in the MLFs [Bi(2.5,\,2.0,\,1.4\,nm)--FeNi(0.8,\,1.2\,nm)]$_{16}$, resulting from the model simulations of the complex dielectric 
response (Eq.\,(1)) (for details see supplementary material).}
\begin{ruledtabular}
\begin{tabular}{lcccccc}
FeNi &Bi &$A_D$&$\gamma_D$&$A_j$&$E_j$& $\gamma_j$\\ 
(nm)   & (nm) &     & (eV)     &    &(eV) & (eV) \\ \hline
    &2.5 &25$\pm$4 &1.4$\pm$0.1  &96$\pm$18  & 0.32  & 0.83\\
1.2 &2.0 &23$\pm$3 &2.2$\pm$0.1  & 97$\pm$9 & 0.40   & 1.08\\
    &1.4 &64$\pm$0.3 &0.75$\pm$0.17& 19$\pm$9 & 0.81 & 1.32\\ \hline
    &2.5 & & & 97$\pm$1 & 0.459 & 1.271 \\
0.8 &2.0 & & & 98$\pm$1 & 0.481 & 1.354 \\
    &1.4 & & & 103$\pm$1 & 0.429& 1.628 \\
\end{tabular}
\end{ruledtabular}
\end{table}
\begin{figure}
\includegraphics*[width=80mm]{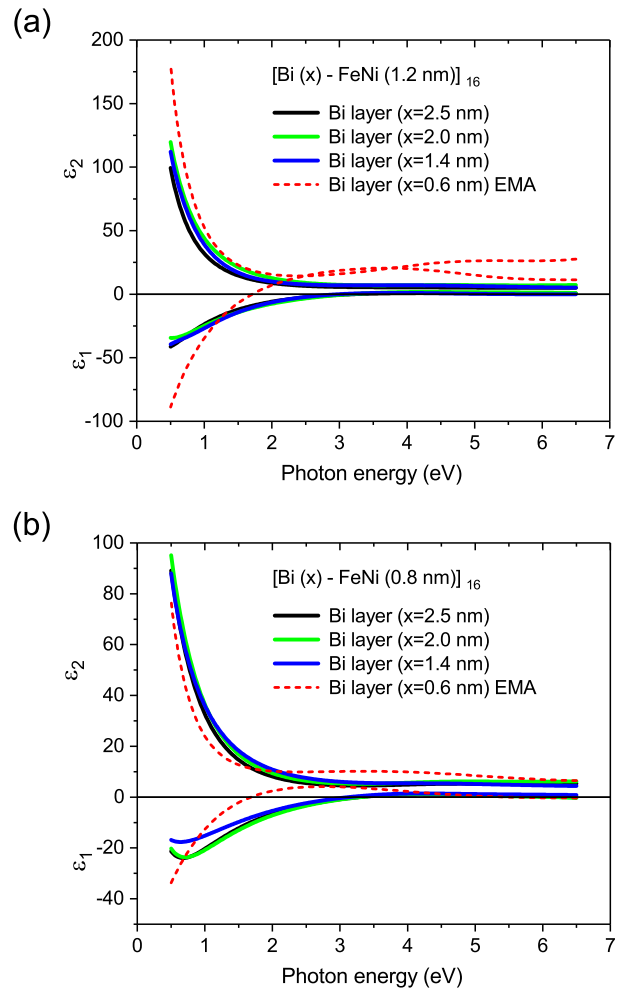}
\caption{The complex dielectric function spectra, $\varepsilon_2(\omega)$ and $\varepsilon_1(\omega)$, of the Bi layer in the 
[Bi(2.5,\,2.0,\,1.4,\,0.6\,nm)--FeNi($h$)]$_{16}$ MLFs for the 
FeNi layer thickness $h$ (a) 1.2\,nm and (b) 0.8\,nm are shown 
by solid black, green, blue, and dashed red curves, respectively. 
\vspace{-0.5cm}}
\label{e1e2}
\end{figure}
The ellipsometric angles $\Psi(\omega)$ and $\Delta(\omega)$ were 
measured for the prepared Al$_2$O$_3$/(Bi-FeNi)$_{16}$/Sitall 
MLF samples at room temperature at three or four angles of incidence 
of 60$^\circ$, 65$^\circ$, 70$^\circ$, and 75$^\circ$ in a wide photon 
energy range of 0.5--6.5\,eV with a J.A. Woollam VUV-VASE 
spectroscopic ellipsometer (Fig.\,\ref{PsiDelta}(a,b) 
illustrates the $\Psi(\omega)$ and $\Delta(\omega)$ measured at 
70$^\circ$). The complex dielectric function $\tilde\varepsilon(\omega)=\varepsilon_1(\omega)+\rm{i} \varepsilon_2(\omega)$ of each Bi or FeNi layer was modeled by 
the Drude term and the sum of Lorentz oscillators to account 
for the contributions of free charge carriers and interband 
optical transitions, respectively   
\begin{eqnarray}
\tilde\varepsilon(E\equiv \hbar\omega)=\epsilon_{\infty}-\frac{A_D}{E^2+{\rm i}E\gamma_D}+\sum_j\frac{A_j \gamma_jE_j}{E_j^2-E^2-{\rm i}E\gamma_j}, 
\label{DispAna}
\end{eqnarray}
where $\epsilon_{\infty}$ is the high frequency dielectric constant. 
The fitted Drude parameters were $A_D$ (related to the plasma 
frequency $\omega_p$ via $A_D=\epsilon_{\infty}\hbar\omega^2_p$) 
and scattering rate $\gamma_D$. The adjustable Lorentz oscillator 
parameters were $E_j$, $\gamma_j$, and $A_j$ of the peak energy, 
full width at half maximum, and $\varepsilon_2$ peak height, respectively. The ellipsometric angles, $\Psi(\omega)$ and $\Delta(\omega)$, 
measured at different angles of incidence were fitted 
simultaneously in the framework of the multilayer model 
Al$_2$O$_3$/[Bi(2.5,\,2.0,\,1.4,\,0.6\,nm)--FeNi(0.8,\,1.2\,nm)]$_{16}$/Sitall, where, in addition, the surface roughness was taken into account 
by the standard effective medium approximation (EMA) based on the 
Bruggeman model (50\% Al$_2$O$_3$ -- 50\% vacuum), using the J.A. Woollam VASE software \cite{WVASE_software}. In the simulation of the 
ellipsometric angles, $\Psi(\omega)$ and $\Delta(\omega)$, 
the Bi layers in each MLF structure were described by the 
dispersion models (Eq.\,(\ref{DispAna})), including three Lorentz 
oscillators and the Drude term where necessary. The discontinuous 
nanoisland FeNi layers were modeled by the effective dielectric 
function in EMA, which describes the optical properties of a complex 
composite by an effective homogeneous medium. 
In the utilized multilayer model, the spectra of the complex dielectric 
function of the blank Sitall substrate obtained from our previous 
SE studies \cite{Kovaleva_metals,Kovaleva_APL_1} were substituted. 
The Bi and FeNi layer thicknesses were fitted to their respective 
nominal values. The good quality of the fit obtained for the 
measured angle of incidence of 70$^\circ$ is demonstrated by Fig.\,\ref{PsiDelta}(a,b), where we plot the recorded ellipsometric angles $\Psi(\omega)$ 
and $\Delta(\omega)$ and the fitting results. 
The details of the used model and the resulting Drude-Lorentz 
parameters along with the fit quality check are given in 
supplementary material to this article.  
The simulation in the framework of the multilayer model for the 
Al$_2$O$_3$/(Bi--FeNi)$_{16}$/Sitall MLFs, where the Bi--FeNi 
interface roughness is explicitly included, does not essentially 
improve the fit (see the analysis presented in \cite{Kovaleva_APL_2}), 
suggesting that the Bi--FeNi interface roughness is essentially 
incorporated in the EMA dielectric function 
of the nanoisland FeNi layers. 

From the multilayer model simulations, the dielectric function 
spectra of the Bi and FeNi layers were obtained 
(see supplementary material for details). 
Here, we are particularly interested in the dielectric function 
spectra, $\varepsilon_1(\omega)$ and $\varepsilon_2(\omega)$, of 
the Bi layer in the studied  
[Bi(2.5,\,2.0,\,1.4,\,0.6\,nm)--FeNi(1.2\,nm)]$_{16}$ and 
[Bi(2.5,\,2.0,\,1.4,\,0.6\,nm)--FeNi(0.8\,nm)]$_{16}$ MLFs, 
and the spectra obtained from the best-fit simulations 
of the $\Psi(\omega)$ and $\Delta(\omega)$ are shown in 
Fig.\,\ref{e1e2}(a,b). One can notice different trends in 
the behavior of the corresponding $\varepsilon_1(\omega)$ 
and $\varepsilon_2(\omega)$ spectra. Thus, 
the $\varepsilon_1(\omega)$ spectra in Fig.\,\ref{e1e2}(b) exhibit 
a clearly pronounced minimum at about 0.8\,eV, 
whereas $\varepsilon_1(\omega)$ spectra in Fig.\,\ref{e1e2}(a) 
display more negative values falling down to --50. Accordingly, 
the $\varepsilon_2(\omega)$ spectra in Fig.\,\ref{e1e2}(a) 
demonstrate a steeper rise at the lowest probed photon energies. 
Moreover, the dielectric function spectra of 
the 0.6\,nm-thick Bi layers in the studied  
[Bi(0.6\,nm)--FeNi(0.8,\,1.2\,nm)]$_{16}$ MLFs display 
apparent trends toward more pronounced metallic behavior at the 
lowest probed photon energies, the $\varepsilon_1(\omega)$ 
exhibits a sharp downturn to negative values, and 
$\varepsilon_2(\omega)$ dramatically increases     
(see Fig.\,\ref{e1e2}(a,b) and supplementary material to this article).
\begin{figure}
\hspace{-0.5cm}\includegraphics*[width=90mm]{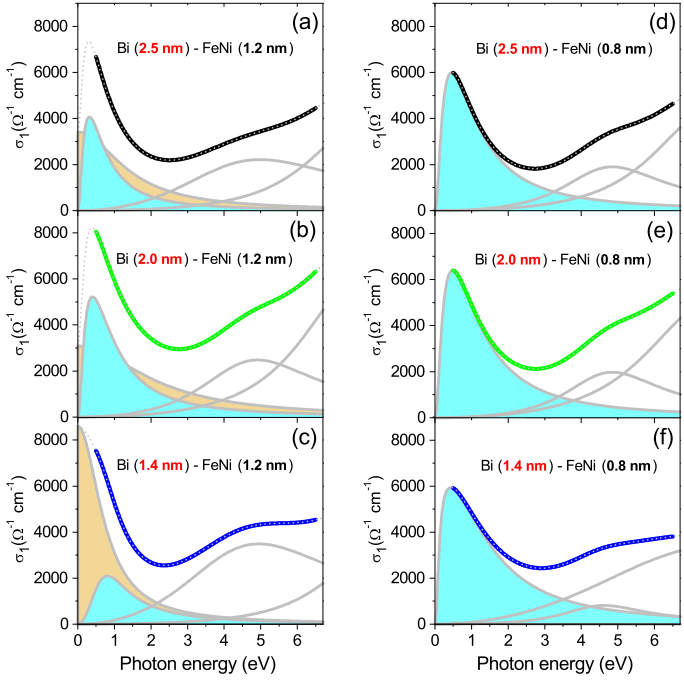}
\caption{The Bi intralayer optical conductivity, 
$\sigma_1(\omega)=\varepsilon_2(\omega)\omega$[cm$^{-1}$]/60, 
in the [Bi--FeNi($h$)]$_{16}$ MLFs 
shown by solid black, green, and blue curves, for the FeNi layer 
thickness $h$ (a--c) 1.2\,nm and (d--f) 0.8\,nm, respectively. 
The contributions from the low-energy Lorentz oscillator and the 
Drude term are indicated by the cyan and yellow shaded area, 
respectively. The summary contribution of the Drude and Lorentz bands 
is shown by dotted lines.
\vspace{-0.5cm}}
\label{Sigma}
\end{figure}

In Fig.\,\ref{Sigma}(a--f) we present the Bi intralayer optical conductivity, $\sigma_1(\omega)=\varepsilon_2(\omega)\omega$[cm$^{-1}$]/60, 
in the studied [Bi(2.5,\,2.0,\,1.4\,nm)--FeNi(0.8,1.2\,nm)]$_{16}$ 
MLFs. Here, the dispersion analysis representation 
resulting from the multilayer model simulations using 
Eq.\,(\ref{DispAna}) is explicitly demonstrated. 
On one hand, the simulation results for the three 
[Bi(2.5,\,2.0,\,1.4\,nm)--FeNi(0.8\,nm)]$_{16}$ MLF structures, 
including the 0.8 nm-thick nanoisland FeNi layer, indicate that 
the Bi layer low-energy response is dominated by a pronounced Lorentz 
band peaking at 0.43\,--\,0.48\,eV having the $\varepsilon_2$ peak 
height of 97\,--\,103 (see Table I). The low-energy interband transition with a high oscillator strength is observed in the dense Bi layers at 
the energy of 0.8\,eV having the $\varepsilon_2$ peak height 
of about 120 \cite{Hunderi,Toudert}. The low-energy peak at 0.8\,eV 
is also seen in the averaged over anisotropy low-energy dielectric 
function response of single crystals \cite{Wang,Lenham}. 
This strong low-energy optical transition is associated with 
interband transitions with the onset near the $\Gamma$ point, 
$\Gamma^+_6-\Gamma^-_6$ and $\Gamma^+_{45}-\Gamma^-_6$ \cite{Golin}, 
and with interband transitions near the T point T$^-_6-$T$^-_{45}$ 
\cite{Liu}. 
Therefore, we can conclude that the low-energy optical response of 
the Bi layer in the [Bi(2.5,\,2.0,\,1.4\,nm)--FeNi(0.8\,nm)]$_{16}$ 
MLFs (see Fig.\,\ref{Sigma}(d--f)) is dominated by the Bi 
semimetallic-like electron band structure.   
On the other hand, our model simulations imply that 
the optical conductivity of the Bi layer in the 
[Bi(2.5,\,2.0,\,1.4\,nm)--FeNi(1.2\,nm)]$_{16}$ 
MLFs, including the 1.2 nm-thick nanoisland FeNi layer, 
has competing contributions from the low-energy Lorentz 
band and from the Drude term 
(see Fig.\,\ref{Sigma}(a--c) and Table I). For the 2.5 and 2.0\,nm 
thick Bi layers, the estimated Drude parameters are 
similar to those characterizing the surface metallic states arising 
due to the Rashba effect in ultrathin Bi(001) films \cite{Yokota} 
($\sigma_{1(\omega\rightarrow 0)}$\,=\,2300\,$\Omega^{-1}\cdot
$cm$^{-1}$ and $\gamma_D$\,=2.0\,eV). 
However, it was shown that the surface layer in the ultrathin 
Bi(012) films can possess a pseudocubic 
Bi$\left\{ 012 \right\}$-oriented allotrope with the even number 
of layers (represented by black phosphorus-like puckered layers) \cite{NagaoPRL}.
We have found that with decreasing the Bi layer thickness from 
2.0 to 1.4\,nm (corresponding to about six and four Bi(012) 
monolayers, respectively), the Drude dc limit 
$\sigma_{1(\omega\rightarrow 0)}$ 
significantly increases from about 3100$\pm$400 to 
8600$\pm$40\,$\Omega^{-1}\cdot$cm$^{-1}$, and the scattering rate 
$\gamma_D$ decreases from 2.2$\pm$0.2 to 0.8$\pm$0.2\,eV. 
At the same time, the low-energy Lorentz band becomes significantly 
suppressed (see Fig.\,\ref{Sigma}(a--c)). 
The observed evolution of the competing Drude and Lorentz parts 
can be attributed to the progressive increase in the 
contribution of the Bi surface metallic states.
However, note the difference from the results obtained for 
the 40--2.8\,nm-thick Bi(001) single-layer films \cite{Yokota}, 
where the $\omega_p$ and $\gamma_D$ increase with a decrease 
of the film thickness.   
We suppose that here a GMR-like case plays an important role. 
Indeed, in the GMR-type MLF structures,  
FM coupling is found for 2.5\,nm and 1.3\,nm-thick spacer 
layers, and AFM coupling is found for a 2\,nm-thick spacer 
layer \cite{Reiss}.
The existence of AFM or FM GMR-type correlations between the 
discontinuous nanoisland FeNi layers could occur in the SFM 
regime \cite{Kovaleva_JNM,Kleemann,Kovaleva_JNR}. 
Therefore, in the studied GMR-type [Bi(2.5,\,2.0,\,1.4\,nm)--FeNi(1.2\,nm)]$_{16}$ MLFs, the magnetic interaction between the neighboring 
FeNi layers, which is responsible for the spin-dependent scattering 
in the magnetic layer, oscillates from FM via AFM to FM ones. 
The spin-dependent scattering at the 
Bi/FeNi interface necessarily affects the scattering of the 
Bi surface metallic charge carriers ($\gamma_D$), 
which should decrease for the FM coupling and increase for 
the AFM coupling between the neighdoring FeNi layers. 
The decrease of the $\gamma_D$ of the surface 
metallic charge carriers in the FM regime will naturally lead to 
the increase in the optical $dc$ conductivity limit 
(see Fig.\,\ref{Sigma}(c)). 
According to the results of the present multilayer model 
simulations for the [Bi(2.5,\,2.0,\,1.4\,nm)--FeNi(0.8,\,nm)]$_{16}$ MLFs, including the 0.8 nm-thick nanoisland FeNi layers, the Bi layer 
exhibits semimetallic bulk-like electron band structure, however, 
the Drude surface metallic conductivity (the Drude term) is implicit 
in Fig.\,\ref{Sigma}(d--f). Here, with decreasing the FeNi layer 
thickness, strong SPM-type fluctuations of giant magnetic moments of the FeNi nanoislands become important \cite{Kovaleva_JNM,Kovaleva_JNR}. 
In our recent study \cite{Kovaleva_srep}, 
we reported that this leads to 
localization phenomena in the GMR-type MLFs, introduced by 
an additional strong magnetic disorder and long-range many-body 
interactions between giant magnetic moments of FeNi nanoislands.  
Therefore, the lack of evidence on the surface 
metallic states for the Bi layer in 
the [Bi(2.5,\,2.0,\,1.4\,nm)--FeNi(0.8,\,nm)]$_{16}$ MLFs 
can be referred to strong SPM-type fluctuations of 
giant magnetic moments of FM FeNi nanoislands, leading to 
strong scattering and localization of scarce free charge carriers in 
the Bi surface layer. In addition, we found that 
the dielectric function spectra of the two 
Bi$\left\{ 012 \right\}$ monolayers demonstrate pronounced 
metallicity properties in the 
[Bi(0.6\,nm)--FeNi(0.8,\,1.2\,nm)]$_{16}$ MLFs.
The origin of the discovered semimetal-to-metall crossover 
needs to be further investigated. In particular, 
the impact of lattice mismatch at the interface on the electron 
band structure of 2D bismuthene \cite{Liu_RSC} 
and the AFM mechanism of a giant SOC-splitting \cite{Rashba1} 
in the presence of AFM spin textures at the interface 
should be considered.  

In conclusion, 
using the advances of the spectroscopic ellipsometry approach, 
we extracted the (pseudo)dielectric function spectra 
of the ultrathin Bi layers incorporating from two to eight 
Bi(012) monolayers in the 
[Bi(0.6,\,1.4,\,2.0,\,2.5,\,nm)--FeNi(0.8,\,1.2\,nm)] 
multilayer structures grown by rf sputtering deposition. 
We found that the Bi(012) layers inside 
the studied multilayer film structures can possess 
the surface metallic conductivity, which is strongly influenced 
by the morphology and magnetic properties of the nanoisland 
FeNi layer. The obtained results may be useful for implementing 
the full potential of the GMR applications based on quasi-2D Bi 
layers.\\

See the supplementary material for the atomic force 
microscopy (AFM), X-ray reflectivity (XRR), and X-ray diffraction 
(XRD) characterization of the MLFs and for details of the 
spectroscopic ellipsometry study.\\

This work was partially supported by the Czech Science Foundation 
(Project No.\,20-21864S), and European Structural and Investment Funds 
and the Czech Ministry of Education, Youth, and Sports 
(Project No.\,SOLID21, CZ.02.1.01/0.0/0.0/16$_{-}$019/0000760). 
The theoretical analysis performed by K.~Kugel was supported by 
the Russian Science Foundation, project No. 21-12-0254 
(https://rscf.ru/en/project/21-12-00254/).\\

The authors have no conflicts to disclose.\\

The data that support the findings of this study are available 
within this article (and its supplementary material).


\begin{thebibliography}{99}

\bibitem{Rashba}Y. A. Bychkov and E. I. Rashba, JETP Lett. 
{\bf 39}, 78 (1984).

\bibitem{Liu_RSC}M.-Y. Liu, Y. Huang, Q.-Y. Chen, Z.-Y. Li, C. Cao, 
and Y. He, 
RSC Adv. {\bf 7}, 39546 (2017).  

\bibitem{Gonze}X. Gonze, J.-P. Michenaud, and J.-P. Vigneron, 
Phys. Rev. B {\bf 41}, 11827 (1990).


\bibitem{Hoffman} C. A. Hoffman, J. R. Meyer, F. J. Bartoli, A. Di Venere, X. J. Yi, C. L. Hou, H. C. Wang, J. B. Ketterson, and G. K. Wong, 
Phys. Rev. B {\bf 48}, 11431 (1993).  

\bibitem{Toudert1} J. Toudert, R. Serna, I. Camps, J. Wojcik, P. Mascher, E. Rebollar, and T. A. Ezquerra, 
J. Phys. Chem. C {\bf 121}, 3511 (2017).

\bibitem{Yokota} K. Yokota, J. Takeda, C. Dang, G. Han, D. N.
McCarthy, T. Nagao, S. Hishita, K. Kitajima, and I. Katayama,  
Appl. Phys. Lett. {\bf 100}, 251605 (2012).

\bibitem{Kovaleva_APL_2} N. N. Kovaleva, D. Chvostova, O. Pacherova, L. Fekete, K. I. Kugel, F. A. Pudonin, and A. Dejneka, Appl. Phys. Lett. {\bf 111}, 183104 (2017).

\bibitem{Kovaleva_srep} N. N. Kovaleva, F. V. Kusmartsev, A. B. Mekhiya, I. N. Trunkin, D. Chvostova, A. B. Davydov, L. N. Oveshnikov, O. Pacherova, I. A. Sherstnev, A. Kusmartseva, K. I. Kugel, A. Dejneka, F. A. Pudonin, Y. Luo, and B. A. Aronzon, Sci. Rep. {\bf 10}, 21172 (2020). 


\bibitem{Kovaleva_JNM} A. Stupakov, A. V. Bagdinov, V. V. Prokhorov, 
A. N. Bagdinova, E. I. Demikhov, A. Dejneka, K. I. Kugel, 
A. A. Gorbatsevich, F. A. Pudonin, \mbox{and N. N. Kovaleva}, 
J.\,Nanomater., 3190260 (2016).

\bibitem{Boltaev} A. P. Boltaev, F. A. Pudonin, I. A. Sherstnev, and D. A. Egorov, 
J. Phys. Condens. Matter {\bf 30}, 295804 (2018). 

\bibitem{Kovaleva_JNR}N. N. Kovaleva, A. V. Bagdinov, A. Stupakov, A. Dejneka, E. I. Demikhov, A. A. Gorbatsevich, F. A. Pudonin, K. I. Kugel, and F. V. Kusmartsev, 
J. Nanopart. Res. {\bf 20}, 109 (2018).

\bibitem{Pudonin_JETP} A. P. Boltaev, F. A. Pudonin, I. A. Sherstnev, and D. A. Egorov, 
J. Exp. Theor. Phys. {\bf 125}, 465 (2017).

\bibitem{Noskova}D. D. Noskova, F. A. Pudonin, I. A. Sherstnev, G. N. Eroshenko, D. A. Egorov, and A. M. Shadrin,
Phys. Lett. A {\bf 410}, 127546 (2021).  

\bibitem{Vedmedenko}E. Y. Vedmedenko, H. P. Oepen, and J. Kirschner,
Phys. Rev. B {\bf 67}, 012409 (2003).

\bibitem{Kleemann}W. Kleemann, O. Petracic, Ch. Binek, G. N. Kakazei, Yu. G. Pogorelov, J. B. Sousa, S. Cardoso, and P. P. Freitas, 
Phys. Rev. B {\bf 63}, 134423 (2001).

\bibitem{WVASE_software} J.A. Woollam {\it VASE Spectroscopic Ellipsometry Data Analysis Software}. (J.A. Woollam Co.: Lincoln, NE, USA, 2010).

\bibitem{Kovaleva_metals}N. Kovaleva, D. Chvostova, and A. Dejneka, 
Metals {\bf 7}, 257 (2017).    

\bibitem{Kovaleva_APL_1} N. N. Kovaleva, D. Chvostova, A. V. Bagdinov, 
M. G. Petrova, E. I. Demikhov, F. A. Pudonin, and A. Dejneka, 
Appl. Phys. Lett. {\bf 106}, 051907 (2015).

\bibitem{Hunderi} O. Hunderi, 
J. Phys. F {\bf 5}, 2214 (1975). 

\bibitem{Toudert} J. Toudert and R. Serna, 
Opt. Mater. Express {\bf 7}, 2299 (2017).  

\bibitem{Wang} P. Y. Wang and A. L. Jain, Phys. Rev. B {\bf 2}, 
2978 (1970). 

\bibitem{Lenham} A. P. Lenham, D. M. Treherne, and R. J. Metcalfe, 
J. Opt. Soc. Am. {\bf 55}, 1072 (1965).

\bibitem{Golin} S. Golin, Phys. Rev. B {\bf 166}, 643 (1968).

\bibitem{Liu} Y. Liu and R. Allen,
Phys. Rev. B {\bf 52}, 1566 (1995).

\bibitem{NagaoPRL}T. Nagao, J. T. Sadowski, M. Saito, S. Yaginuma, Y. Fujikawa, T. Kogure, T. Ohno, Y. Hasegawa, S. Hasegawa, and T. Sakurai, 
Phys. Rev. Lett. {\bf 93}, 105501 (2004).

\bibitem{Reiss}A, H\"utten, S. Mrozek, S. Heitmann, T. Hempel, 
H. Br\"uckl, and G. Reiss,
Acta Mater. {\bf 47}, 4245 (1999).

\bibitem{Rashba1}L.-D. Yuan, Z. Wang, J.-W. Luo, E. I. Rashba, and A. Zunger,
Phys. Rev. B {\bf 102}, 014422 (2020).

















\end{thebibliography}
\end{document}